\documentclass[paper=a4, fontsize=12pt]{scrartcl}
\usepackage[]{algorithm2e}
\usepackage[english]{babel}															
\usepackage{amsmath,amsfonts,amsthm} 
\usepackage{url}
\usepackage{algpseudocode}
\usepackage{algorithmicx}
\usepackage[breaklinks=true]{hyperref}
\usepackage{breakcites}
\usepackage{contour}
\usepackage{amsmath}
\usepackage{bbm}

\theoremstyle{definition}

\theoremstyle{remark}

\makeatletter
\let\c@equation\c@thm
\makeatother
\numberwithin{equation}{section}

\usepackage{sectsty}
\allsectionsfont{\centering \normalfont\scshape}

\usepackage{fancyhdr}
\numberwithin{equation}{section}		
\numberwithin{figure}{section}			
\numberwithin{table}{section}				

\newcommand{\horrule}[1]{\rule{\linewidth}{#1}} 	

\title{
		\usefont{OT1}{bch}{b}{n}
		\normalfont \normalsize \textsc{University Joseph Fourier, Grenoble-I} \\ 
		\horrule{1pt} \\
		\Large Certification of programs with computational effects  \\
		\horrule{1pt} \\
}
\author{
		\normalfont 								\normalsize
        Burak Ekici\\[-5pt]		\normalsize
        May 12, 2014
}
\date{}
\begin{document}
\small
\maketitle
\begin{abstract}

 \textbf{\textsc{Abstract}}. In purely functional programming languages imperative features, more generally computational effects are prohibited. However, non-functional languages do involve effects. The theory of decorated logic provides a rigorous formalism (with a refinement in operation signatures) for proving program properties with respect to computational effects. The aim of this thesis is to first develop Coq libraries and tools for verifying program properties in decorated settings associated with several effects: states, local state, exceptions, non-termination, etc. Then, these tools will be combined to deal with several effects. 
\end{abstract}

%

The syntax of any programming language defines the set of rules that a programmer may utilize during coding progress while its semantics stand for proving program properties. In a purely functional programming language, a term $f$ with an argument of type $X$ and a result of type $Y$ (which may be written $f: X \to$ Y)  is denotationally interpreted as a function $f$ between the sets $[\!\![$X$]\!\!]$ and $[\!\![$Y$]\!\!]$. It follows that, when an operation has several
arguments, they can be evaluated in parallel, or
in any order. It is possible to interpret a purely functional
programming language via categorical semantics based on
\emph{cartesian closed categories}. The word \emph{``cartesian``} here
refers to the categorical
products, which are interpreted as
cartesian products of sets and used for dealing
with pairs of arguments. The
logical semantics
of such a language defines a set of rules that may be used for
proving properties of programs.
But non-functional programming languages such as IMP, 
C or Java do include computational effects. For instance, a
C or IMP function may modify the state structure and a
Java function may throw an exception during the computation. Such
operations are examples of computational effects. To cope with them,~\cite{DDR11} provides a canonical approach. It indeed uses \emph{\textbf{decorations}} on the main operators of the effect  (\emph{with superscripts between parentheses}) instead of mentioning the effect itself. I.e., any \emph{state accessor function} $f$, seen as $X \times S \to Y$ can be interpreted as $f^{(1)}: X \to Y$ where $X$ and $Y$ are sets with the distinguished set of states $S$ and with \emph{cartesian product operator}, $``\times``$. Thus, this approach proposes a refinement in the operator signatures by keeping them closer to syntax where there is no effect appearance but instead \emph{\textbf{decorations}}. Technical details of the refinement are based on \emph{category theoretical objects}: see procedure~\ref{algo:acc}~.  

Accordingly, my PhD study simply focuses on formal models of computational effects (with effect combinations) through \emph{decorated logic} over \emph{cartesian effect categories}~\cite{DDR11}. Besides, we develop \href{http://coq.inria.fr/}{Coq} libraries to verify program properties with respect to effects in question. 

\begin{algorithm}
 \begin{algorithmic}[1]
\small
  \label{algo:acc}
 \Procedure{Dec. acc.}{$\mathbb{C}, \texttt{S}$} Cartesian effect category $\mathbb{C}$, distinguished states object $\texttt{S}$ 
\DontPrintSemicolon 
\State $\mathtt{\Phi}\colon \mathbb{C} \to \mathbb{C}$ is an endo-functor defined as follows:{\;
\ \ \ \ $\mathtt{\Phi}(\mathtt{X}) = \mathtt{X} \times \mathtt{S}$ 
\ \ \  and \ \ \ $\mathtt{\Phi}(\mathtt{f}\colon \mathtt{X} \to \mathtt{Y}$) = ($\mathtt{f} \times \mathtt{id_S})\colon \mathtt{X} \times \mathtt{S} \to \mathtt{Y} \times \mathtt{S}$ \;
}
\State $\mathtt{cM}$ ($\mathtt{\Phi}, \mathtt{\delta}\colon \mathtt{\Phi} \Rightarrow \mathtt{\Phi^2}, \mathtt{\epsilon}\colon \mathtt{\Phi} \Rightarrow \mathtt{id_{\mathbb{C}}} $) is the $\texttt{states comonad}$ with following settings:
{\;
\ \ \ \ $\mathtt{\delta}_{X} \colon \mathtt{X} \times \mathtt{S} \to \mathtt{X} \times \mathtt{S} \times \mathtt{S}$ \ \ \ 
and \ \ \ $\mathtt{\epsilon}_{X}\colon \mathtt{X} \times \mathtt{S} \to \mathtt{X}$
}
{\;
\ \ \ \ \ \ \ \ \ \ $(x, s)  \mapsto (x, s, s)$ \ \ \ 
\ \ and \ \ \ \ \ \ \ \ \  $(x, s) \mapsto x$ \;
}
\State$\mathbb{C}_1$ is the $\mathtt{coKleisli}$ category of $\mathtt{cM}$ over $\mathbb{C}$ defined as: {\;
\ \ \ \  $\mathtt{Obj ({\mathbb{C}_1})} = \mathtt{Obj ({\mathbb{C}})}$, \ \  $\mathtt{Hom_{\mathbb{(C)}}(X, Y)} = \mathtt{Hom_{({\mathbb{C}_1)}}(\mathtt{\Phi}X, Y)}$ $g^{(1)} \circ_{\mathbb{C}_1} f^{(1)} = g_0 \circ_{\mathbb{C}} \Phi(f_0) \circ_{\mathbb{C}} \delta$
\;
}
\State \Return{Any accessor $\mathtt{f_0}\colon \mathtt{X\times S} \to \mathtt{Y} \in \mathtt{Hom_{\mathbb{(C)}}}$ is interpreted as $\mathtt{f^{(1)}}\colon \mathtt{X} \to \mathtt{Y} \in \mathtt{Hom_{(\mathbb{C}_1)}}$}

\EndProcedure
\end{algorithmic} 
\end{algorithm}

Following is the brief description to the procedure~\ref{algo:acc}~: the \emph{endofunctor}, ${\Phi}$ on \emph{cartesian effect category of \textbf{sets}}, $\mathbb{C}$, is a \emph{comonad} with \emph{natural transformations} $\mathbb{\delta}$ and $\mathbb{\epsilon}$. Thus, this triple yields in the existence of the \emph{coKleisli category}, $\mathbb{C}_1$. Therefore, any \emph{impure} function $f^{(1)}: X\to Y \in \mathbb{C}_1$ is the interpretation of $f_{0}: X\times S \to Y \in \mathbb{C}$ representing \emph{accessors}.
 
To interpret \emph{modifiers}, the \emph{endofunctor}, ${\Phi_1}$(having the same settings with $\Phi$) on category, $\mathbb{C}_1$ with compatible \emph{natural transformations} $\mathbb{\mu}$ and $\mathbb{\eta}$ is proven as a \emph{monad} on which a \emph{Kleisli category}, $\mathbb{C}_2$ is built. Similarly, any \emph{impure} function $f^{(2)}: X\to Y \in \mathbb{C}_2$ is the interpretation of $f_{0}: X\times S \to Y\times S \in \mathbb{C}$ thus representing \emph{modifiers}. Additional to those syntactical tricks, we also have an \emph{algebra} for \emph{states effect} mainly based on \emph{monadic equational logic}, \emph{categorical products} and some 
\emph{observational properties}. The \emph{hierarchy rules} define the transition between different sorts of operators: a \emph{pure function} can be seen as an \emph{accessor} and similarly \emph{an accessor} function can be seen as a \emph{modifier}. For details, see~\cite{DBLP:journals/corr/abs-1112-2396}. 

A generic \href{https://forge.imag.fr/frs/download.php/522/STATES-0.6.tar.gz}{\emph{Coq library}} formalizing the \emph{states effect} in above given settings was developed and detailed in~\cite{DDEP14}. The main idea here is to prove  the 7 primitive properties of the states structure proposed by~\cite{PP02}. This provides an environment in which programmers are enabled to prove program properties through already proven lemmas with respect to \emph{states effect}. Those proofs become crucial when the order of evaluation is not specified or more generally when parallelization comes into play~\cite{LG88}. To check the \emph{soundness} of this proof system, we have specialized the generic library for the case of IMP language and proven equalities between some IMP programs involving \emph{terminating loops} and \emph{conditionals}. See the \href{https://forge.imag.fr/frs/download.php/569/IMP-STATES-0.1.tar.gz}{link}~.

Thanks to the \emph{duality} between \emph{exceptions} and \emph{states}~\cite{DDFR12},
all the syntactical tricks together with the algebra for states have been dualized for exceptions. I.e., \emph{lookup} operation on the state is dual to \emph{tagging} an exception. Thus, an environment to cope with \emph{exceptions effect} has been ensued and developed in \texttt{Coq}. Here is the \href{https://forge.imag.fr/frs/download.php/541/EXCEPTIONS-0.1.tar.gz}{link} to the generic library for \emph{exceptions}. In addition, we have combined mentioned effects through the following non-canonical way: just composing \emph{functors} and merging the \emph{algebras}. In order to see its \emph{soundness}, we have  specialized the library for \{IMP+Exc\}\footnote{The fact that IMP has no \emph{exception handling mechanism}, we simply defined \emph{throw} operation and \emph{try-catch} block as additional commands. The resulting language is called \{IMP+Exc\}. It is developed in \texttt{Coq} syntax but in IMP semantics where all commands have $\mathbbm{1}\to \mathbbm{1}$ type.} language. It can be found through the \href{https://forge.imag.fr/frs/download.php/557/IMP-STATES-EXCEPTIONS-0.1.tar.gz}{link}~. It apparently follows that in this library \emph{equational proofs} between  \{IMP+Exc\} programs can be stated including both \emph{states} and \emph{exceptions effects}.

Considering future directions, we are planning to introduce a \emph{canonical framework} first to combine \emph{states} and \emph{exceptions} in \emph{decorated settings}. Then, we will make attempts to generalize the idea to the other ones. In the mean time, the comparison with \emph{monad transformers} is planned to be stated.


\begin{thebibliography}{99} 
\bibitem[Dumas:2014:coqstates]{DDEP14}
Jean-Guillaume Dumas and Dominique Duval and Burak Ekici and Damien Pous.
\newblock Formal verification in {Coq} of program properties involving the global state effect. 
\newblock In JFLA, Fr\'ejus, 2014.


\bibitem[Dumas:2012:states]{DBLP:journals/corr/abs-1112-2396}
Jean-Guillaume Dumas, Dominique Duval, Laurent Fousse, and Jean-Claude Reynaud.
\newblock Decorated proofs for computational effects: States.
\newblock In {\em ACCAT}, pages 45--59, 2012.


\bibitem[Dumas:2011a]{DDR11} 
Jean-Guillaume Dumas, Dominique Duval, Jean-Claude Reynaud.
Cartesian effect categories are Freyd-categories.
Journal of Symbolic Computation 46, p.~272-293 (2011).

\bibitem[Dumas:2012:duality]{DDFR12}
Jean-Guillaume Dumas, Dominique Duval, Laurent Fousse and Jean-Claude Reynaud.
Journal of Mathematical Structures in Computer Science 22, p.~719-722(2012).


%
%
%

\bibitem[Plotkin\&Power:2002]{PP02} 
Gordon D. Plotkin, John Power.
Notions of Computation Determine Monads. 
FoSSaCS 2002.
Springer-Verlag 
Lecture Notes in Computer Science 2303, p.~342-356 (2002).

\bibitem[Lucassen\&Gifford:1988]{LG88}
J.~M. Lucassen and D.~K. Gifford.
\newblock Polymorphic effect systems.
\newblock In J.~Ferrante and P.~Mager, editors, \emph{POPL}, pages 47--57. ACM,
  1988.
\newblock ISBN 0-89791-252-7.

\end{thebibliography}
\end{document}